\documentclass[runningheads]{llncs}
\usepackage{graphicx}
\usepackage{amsmath,amssymb} % define this before the line numbering.

\usepackage{color}
\usepackage[width=122mm,left=12mm,paperwidth=146mm,height=193mm,top=12mm,paperheight=217mm]{geometry}
\usepackage{caption}
\usepackage{subcaption}
\usepackage{siunitx}
\usepackage[inline]{enumitem}
\usepackage{booktabs}

\makeatletter
\def\blfootnote{\xdef\@thefnmark{}\@footnotetext}
\makeatother

\usepackage{tikz}
\usepackage{pgfplots}

\begin{document}
%\renewcommand\thelinenumber{\color[rgb]{0.2,0.5,0.8}\normalfont\sffamily\scriptsize\arabic{linenumber}\color[rgb]{0,0,0}}
%\renewcommand\makeLineNumber {\hss\thelinenumber\ \hspace{6mm} \rlap{\hskip\textwidth\ \hspace{6.5mm}\thelinenumber}}
%\linenumbers
\pagestyle{headings}
\mainmatter
\def\ECCV20SubNumber{4571}  % Insert your submission number here

\title{Practical Deep Raw Image Denoising\\ on Mobile Devices}
%\title{SuperIQ: AI-Enhanced Night Shot for Smartphones} % Replace with your title
% \titlerunning{ECCV-20 submission ID \ECCV20SubNumber}
% \authorrunning{ECCV-20 submission ID \ECCV20SubNumber}

\author{
	Yuzhi Wang\inst{1,2}%
	\and Haibin Huang\inst{2}%
	\and Qin Xu\inst{2}%
	\and Jiaming Liu\inst{2}%
	\and Yiqun Liu\inst{1}%
	\and Jue Wang\inst{2}%
}
\authorrunning{Y. Wang, H. Huang, Q. Xu, J. Liu, Y. Liu, J. Wang}
\institute{
	Tsinghua University \and Megvii Technology
}

\maketitle

\blfootnote{{This work is supported by The National Key Research and Development Program of China under Grant 2018YFC0831700.}}

\begin{abstract}
%In this work, we present a simple, efficient and general framework for image denoising on mobile devices. Specifically, we propose a learning based approach that can efficiently run on mobile devices with high quality noise reducing. Unlike previous methods, our method doesn't require complicate network architectures but an accurate modeling over the sensor noise property. We first develop a simple noise parameter estimation method by collecting raw images taken under specified light conditions. We further unify the noise parameters of different ISOs into a single noise model. The noise model with estimated parameters then can be applied to noise-free raw images to generate synthetic noise datasets and simulate various inputs. Due to simplification of our noise modeling, we can train a small enough network that can capture a large range of noise and still efficiently run on mobile devices. Extensive experiments were conducted to demonstrate the effectiveness of our method which achieves a good balance between quality and speed.
Deep learning-based image denoising approaches have been extensively studied in recent years, prevailing in many public benchmark datasets. However, the stat-of-the-art networks are computationally too expensive to be directly applied on mobile devices. In this work, we propose a light-weight, efficient neural network-based raw image denoiser that runs smoothly on mainstream mobile devices, and produces high quality denoising results. Our key insights are twofold:
% (1) by leveraging sensor noise calibration, a smaller network that is trained on sensor-specific data can out-perform larger ones trained on general data; 
(1) by measuring and estimating sensor noise level, a smaller network trained on synthetic 
sensor-specific data can out-perform larger ones trained on general data;
(2) the large noise level variation under different ISO settings can be removed by a novel {\em k-Sigma Transform}, allowing a small network to efficiently handle a wide range of noise levels.
We conduct extensive experiments to demonstrate the efficiency and accuracy of our approach. 
Our proposed mobile-friendly denoising model runs at $\sim$70 milliseconds per megapixel on Qualcomm Snapdragon 855 chipset, and it is the basis of the night shot feature of several flagship smartphones released in 2019. 
\end{abstract}

\section{Introduction}

\begin{figure}[htpb]
	\centering
	\begin{center}
		\includegraphics[width=\linewidth]{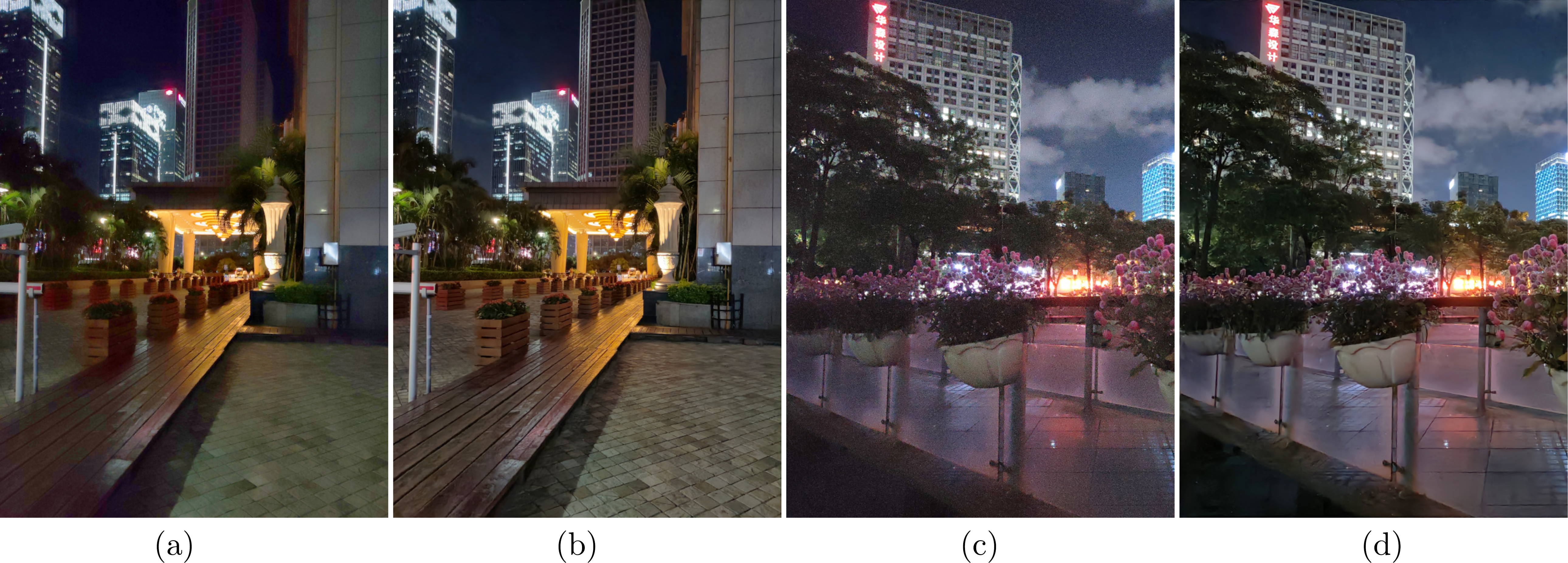}
	\end{center}
	\caption{Our proposed denoising method can run smoothly on smartphones with high quality noise reduction even in low-light conditions (see (b),(d)). In contrast, the default ISP image denoiser produces images with over-smoothed high texture regions (e.g. ground in (a)) and noisy smooth regions (e.g. sky in (c)).} 
	\label{fig:teaser}
\end{figure}

Smartphones have become the go-to devices for consumer photography in recent years. Compared with DSLR cameras, images captured with mobile devices are more easily contaminated with higher level of noise due to the use of relatively low-cost sensors and lenses, especially in low-light scenarios. 

Despite decades of development in image denoising technologies, it remains challenging to restore high quality images from extremely noisy ones on mobile devices. %Traditional methods ~\cite{zhou2018survey,wang2019video,dabov2008image,gu2014weighted,xu2017multi,yair2018multi} utilize natural image statistics hand-crafted features with complex rules to tackle this problem. Such empirical methods would suffer from limited performance in general scenes.  
Recently, deep neural network~(DNN) based denoising methods~\cite{tai2017memnet,chen2017trainable,zhou2019awgn,jain2009natural,xie2012image,mao2016image,ulyanov2018deep} have achieved tremendous success and outperformed most traditional methods~\cite{zhou2018survey,dabov2008image,gu2014weighted,xu2017multi,yair2018multi}.  It is however not practical to directly deploy 
these heavy-weight DNNs on mobile devices due to the limited computational resources available on them. 

In this work, we propose a simple yet efficient approach for deep raw image denoising. It can run efficiently on off-the-shelf smartphones with high quality noise reduction. Our key observation is that the noise characteristics for a specific sensor model are consistent and can be measured with sufficient accuracy. By capturing and modeling sensor noise, we can generate synthetic datasets with clean and noisy image pairs, and train a light-weight neural network on them. The trained model remains highly effective on real images captured by the same sensor (i.e. the same smartphone model). Furthermore, based on the parametric sensor noise model, we derive a unique linear transform in luminance space, dubbed as k-Sigma Transform, that maps noisy images captured under different ISO settings into an ISO-invariant signal-noise space, allowing a single network to handle different noise levels in different scenes. We show that this approach is not only theoretically elegant, but in practice is more efficient than training a separate model for each ISO setting, or having one large model trained on images with different noise levels.  

%Our previous denosing neural netowrks, our method doesn't require complicate network architectures but an accurate modeling over the sensor noise property. We first develop a simple noise parameter estimation method by collecting raw images taken under specified light conditions. We further unify the noise parameters of different ISOs into a single noise model. The noise model with estimated parameters then can be applied to noise-free raw images to generate synthetic noise datasets and simulate various inputs. Due to simplification of our noise modeling, we can train a small enough network that can capture a large range of noise and still efficiently run on mobile devices.

To summarize, the main contributions of this work are as follows:
\begin{itemize}
\item A systematic approach to estimate sensor noise and train a sensor-specific denoising neural network using properly constructed synthetic data.

\item A novel k-Sigma Transform to map noisy images under different ISO settings into a ISO-invariant signal-noise space. Instead of training separate models per ISO or a larger model to cover the variations, the proposed  transform allows a single small network trained in this space to handle images with different noise levels. 

\item A mobile-friendly network architecture for efficient image denoising.  We provide in-depth analysis and comparison with different network architectures and denoising methods, demonstrating that our method has compatible performance with state-of-the-art approaches with significantly less computational resources.      

\end{itemize}

To the best of our knowledge, our solution is the first practical deep-learning-based image denoising approach that has satisfactory efficiency and accuracy on mobile devices. In Fig.~\ref{fig:teaser} we show examplar images captured by an off-the-shelf smartphone that use our approach in low-light photography. Compared with the default ISP image denoising technique, our results contain much more fine details of the scene.  

%1. it is hard to apply denosing for mobile device, with good performance and reasonable speed. 

%2. traditional methods can not handle input with variations in lights, sensor etc. deep learning methods works better but has limitations both in quality and speed.

%3, to have better performance, we focus the problem in raw noise domain. where we can estimate the noise property accurately. Instead of tuning network structure, we can use a much smaller network capture the noise model. 

%4. we further unify the noise model to cover different ISO range, such that a single model can handle different inputs.

%5 experiments

\section{Related work}

%\subsection{Image Denoising}
Image denoising is a fundamental task in image processing and computer vision.
Classical methods often rely on using sparse image priors, such as non-local means (NLM) \cite{buades2005non}, sparse coding \cite{elad2006image,mairal2009non,aharon2006k}, 3D transform-domain filtering (BM3D) \cite{dabov2008image}, and others \cite{gu2014weighted,portilla2003image}. Among them BM3D is usually deemed as the leading method considering its accuracy and robustness. Most of these methods are designed for general noise and do not take advantage of known sensor noise characteristics. Their algorithmic complexity is usually high, making full-fledged implementation difficult on smartphones. 

With the recent development of convolutional neural networks (CNNs), training end-to-end denoising CNNs has gained considerable attention. Earlier work that uses multi-layer perceptron (MLP) \cite{burger2012image} has achieved comparable results with BM3D. Further improvements have been achieved with the introduction of more  advanced network architectures, resulting in a large number of CNN-based denoising methods \cite{tai2017memnet,chen2017trainable,jain2009natural,xie2012image,mao2016image,ulyanov2018deep,zhang2018ffdnet,lehtinen2018noise2noise}.
These works are primarily focused on novel network structures for improving the accuracy, without paying much attention to their adaptability to mobile devices. 

Our work focuses on denoising raw image, i.e., images read out from the sensor in the raw Bayer format before demosaicing and other ISP-processing. On the recently proposed public raw image denoising benchmark datasets \cite{SIDD_2018_CVPR,chen2018learning,anaya2018renoir},  CNN-based methods \cite{chen2018learning,hirakawa2006joint,gharbi2016deep} have achieved the best results. 
It is however a very tedious work to construct such high quality real datasets with clean and noisy image pairs. 
Thus, the problem of synthesizing realistic image noise for training has also been extensively studied, including Gaussian-Poisson noise\cite{foi2008practical,liu2014practical}, Gaussian Mixture Model (GMM) \cite{zhu2016noise}, in-camera process simulation \cite{liu2008automatic,guo2018toward}, GAN-generated noises \cite{chen2018image} and so on. It has been shown that networks properly trained from the synthetic data can generalize well to real data \cite{zhou2019awgn,brooks2019unprocessing}.

The existing best practice for raw image denoising on mobile devices is to capture and merge multiple frames \cite{Hasinoff2016,Liba2019,Mildenhall2017}.
These methods generally require accurate and fast image alignment, which is hard to achieve when moving objects present in the scene. Furthermore, when noise level is high, averaging multiple frames can reduce, but not completely remove the noise, leading to unsatisfactory results.  
To the best of our knowledge, our proposed method is the first single frame, deep learning-based raw image denoiser specifically designed for mobile devices.

\section{Method}
\label{sec:method}

In this section, we first revisit the general ISO-dependent noise model of camera sensor, and then describe how to estimate noise parameters given a new sensor. We further show how to synthesize ISO-independent training data using the proposed k-Sigma Transform, and use it to train a small neural network that can handle a wide range of noise levels. 

\begin{figure}[htpb]
	\centering
	\begin{center}
		\includegraphics[width=.6\linewidth]{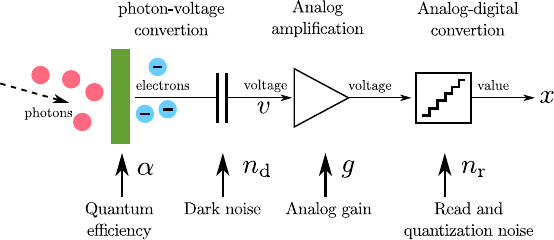}
	\end{center}
	\caption{Photon transfer pipeline: multiple noise sources like shot noise, read-out noise and thermal noise are involved along the camera pipeline. Check \cite{EMVA2010} for more details.}
	\label{fig:pipeline}
\end{figure}

\subsection{The Noise Model}
%Noise modeling has been an active research topic and several noise models have been widely used in previous work, such as the Additive White Gaussian Noise (AWGN) with the homoscedastic Gaussian assumption~\cite{}. Considering the raw pixel readout from sensor is affected by noise in multiple processing stages, it is more accurate to assume a Poisson-Gaussian model ~\cite{foi2008practical,liu2014practical}, where the noise is a combination of a signal-dependent Poisson distribution and a signal-independent Gaussian distribution.  
A camera sensor converts the photons hitting the pixel area during the exposure time into a digitized luminance map. As shown in the photon transfer pipeline shown in Fig.~\ref{fig:pipeline}, this process contains multiple stages, where each stage introduces specific noise. Let us first consider an ideal system with no noise. Under the linear camera model, at each pixel, the sensor conversion is a linear amplification as:
\begin{equation}
   x^* = g\alpha u^*,
\label{eq:noise_photon_transfer}
\end{equation}
where $u^*$ is the expected number of photons hitting the pixel area, $\alpha$ is the quantum efficiency factor and
$g$ is the analog gain. Now considering the system noise in each step of the pipeline in Fig.~\ref{fig:pipeline}, we have:
\begin{equation}
	x = g (\alpha u + n_\text{d})  + n_\text{r},
\label{eq:noise_gain}
\end{equation}
where $u$ denotes the actual collected amount of photons, and $n_d\sim \mathcal{N}(0, \sigma_d^2)$ and  $n_r \sim \mathcal{N}(0, \sigma_r^2)$ are Gaussian noise before and after applying the analog gain. Furthermore, 
it is demonstrated in~\cite{EMVA2010} that $u$ obeys a Poisson distribution of $u^*$, given by

\begin{equation}
    u \sim \mathcal{P}(u^*).
\label{eq:noise_poisson}
\end{equation}
%The relationship between an observed image $I$ and its underlying noise-free image $I^*$ is modeled as 
%\begin{equation}
%    I = I^* + n \sim \mathcal{P}(I^*),
%\label{eq:noise}
%\end{equation}
%where $n$ is the image noise. Modeling $n$ has been an active research topic and several noise models have been widely used in previous work, such as the Additive White Gaussian Noise (AWGN) with the homoscedastic Gaussian assumption~\cite{}. 
%Considering the raw pixel readout from sensor is affected by noise in multiple processing stages, it is more accurate to assume a Poisson-Gaussian model ~\cite{foi2008practical,liu2014practical}, where the noise is a combination of a signal-dependent Poisson distribution and a signal-independent Gaussian distribution:

%\begin{equation}
%    x \sim   P( x^*)  + \mathcal{N}(0, \sigma^2 ),
%\end{equation}
%where $x$ is the pixel readout and $x^*$ is the true signal value.

 %where $x_i$ is the pixel readout and $x^*_i$ is the true signal value. $v_{i} \sim \alpha P(\frac{v_i^*} {\alpha})$ is the sensor noise, and $n_{i} \sim \mathcal{N}(0, \sigma_i^2)$ is reading noise and $g$ as the analog gain
 %we can get the Poisson-Gaussian model of noise as in Equation ~\ref{eq:noise_all}.   

Combining Eqn.~\eqref{eq:noise_photon_transfer} to Eqn.~\eqref{eq:noise_poisson}, we have:
\begin{equation}
    x \sim  (g \alpha ) \mathcal{P}( \frac{x^*} {g \alpha})  + \mathcal{N}(0, g^2\sigma_\text{d}^2 + \sigma_\text{r}^2 ).
\label{eq:noise_all}
\end{equation}
This is consistent with the Poisson-Gaussian noise model that has been extensively studied in previous work~\cite{foi2008practical,liu2014practical}.
This formulation can be further simplified by replacing $k =  g\alpha$ and $\sigma^2 = g^2\sigma_d^2 + \sigma_r^2$:
\begin{equation}
    x \sim  k\mathcal{P}( \frac{x^*} {k})  + \mathcal{N}(0,  \sigma^2 ).
\label{eq:noise_simple}
\end{equation}
Note that both $k$ and $\sigma^2$ are related to $g$, which is determined by the ISO setting of the camera.

\subsection{Parameter Estimation}
\label{sec:method:param}

To sample the distribution described in Eqn.~\eqref{eq:noise_simple}, we need an accurate estimation of $k$ and $\sigma$ under a specified ISO setting of a specific sensor. Luckily, as we check the mean and variance over $x$, shown in Eqn.~\eqref{eq:p_mean_variance}, we can turn it into the following linear regression problem:% by fitting intercept and slope.

%these two variables control different parts of the distribution, allowing us to estimate them separately under different experiment setups. 

%photon sensing,
%\paragraph{Gaussian component:} since the Gaussian noise is signal-independent, we can measure its variance by taking sample images in a complete dark environment. We then use all pixels in these sample images to estimate $\sigma$.

%\paragraph{Poissonian component:} Given the properties of Poisson distribution, from Eqn.~\ref{eq:noise_all} we have:

\begin{equation}
\left\{\begin{array}{ll}
	\mathrm{E}(x) &= x^*, \\
    \mathrm{Var}(x) &= kx^* + \sigma^2.
    \end{array}\right.
\label{eq:p_mean_variance}
\end{equation}

\begin{figure}[tpb]
	\centering
	\includegraphics[width=\linewidth]{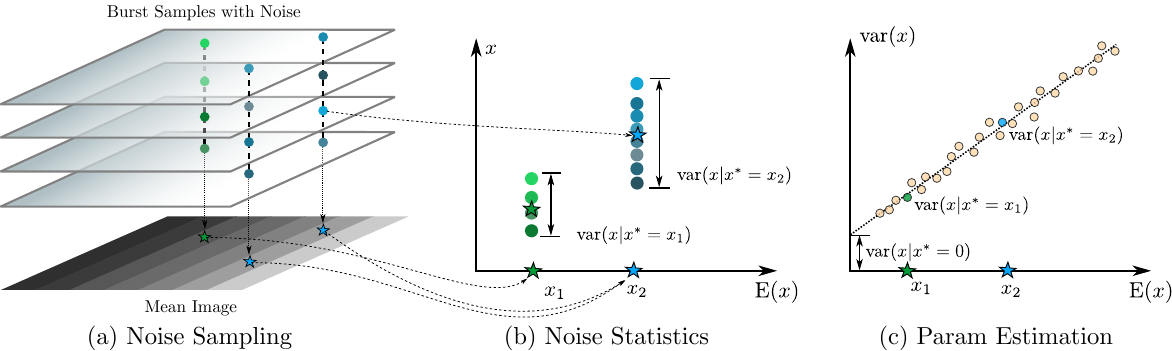}
	\caption{Noise parameter estimation with a burst series of raw images of a static grayscale chart.}%
	\label{fig:estimation}
\end{figure}

%Parameter $k$ thus can be computed as $(Var(x) - \sigma^2)/E(x)$. 
Similar to~\cite{foi2007noise},  we capture a series of raw images of a static grayscale chart in burst mode, depicted in Fig.~\ref{fig:estimation}a, and compute $\mathrm{E}(x)$ from the series of luminance values at the same pixel location. Next, as shown in Fig.~\ref{fig:estimation}b, we bracket all pixels that have the same estimated luminance, and compute $\mathrm{Var}(x)$ from them. A linear regression is then applied to find the optimal estimation of $k$ and $\sigma^2$, illustrated in Fig.~\ref{fig:estimation}c.

%For variance estimation of specified value,  we can look up all valid values with a pre-defined threshold and compute the variance and minus $\sigma^2$ from previous Gaussion part.

\subsection{The k-Sigma Transform}

In real applications the camera will automatically adjust the ISO settings according to the scene illumination, thus one has to consider different noise levels when training the denoising neural network. A straightforward solution is to train a single network to cover a wide range of ISO settings, but it puts extra burden on the network itself as the noise variation in the training data becomes quite large. Inspired by variance stabilizing transformations~\cite{anscombe1948transformation,makitalo2010optimal}, here we propose a k-Sigma Transform to avoid this problem. 

Specifically, we define a linear transform
\begin{equation}
    f(x) = \frac{x}{k} + \frac{\sigma^2}{k^2}.
\label{eq:noise_k_sigma_transfor}
\end{equation}

According to our noise model of Eqn.~\eqref{eq:noise_simple},
\begin{equation}
    f(x) \sim \mathcal{P}( \frac{x^*} {k})  + \mathcal{N}(\frac{\sigma^2}{k^2},  \frac{\sigma^2}{k^2}).
\label{eq:f_x_dist1}
\end{equation}

% \begin{figure}[htpb]
% 	\centering
% 	\includegraphics[width=.7\linewidth]{figs/dist_comp.pdf}
% 	\caption{The value of $\frac{\sigma^2}{k^2}$ under different ISOs measured on three phones with different sensors. For most ISO settings $\frac{\sigma^2}{k^2}$ is sufficiently large. See text for detailed explanation.} %
% 	\label{fig:k_sigma}
% \end{figure}

To analyze this distribution, a usual simplification is to treat the Poisson distribution $\mathcal{P}(\lambda)$
as a Gaussian distribution of $\mathcal{N}(\lambda, \lambda)$~\cite{foi2008practical}. Therefore:
\begin{equation}
	\begin{aligned}
		& P( \frac{x^*} {k}) + \mathcal{N}(\frac{\sigma^2}{k^2},  \frac{\sigma^2}{k^2}) \\
		\approx & \mathcal{N}(\frac{x^*} {k},\frac{x^*} {k})  + \mathcal{N}(\frac{\sigma^2}{k^2},\frac{\sigma^2}{k^2}) \\
		= & \mathcal{N}(\frac{x^*}{k} + \frac{\sigma^2}{k^2},\frac{x^*}{k} + \frac{\sigma^2}{k^2}) \\
    	= & \mathcal{N}[f(x^*), f(x^*)].
	\end{aligned}
    \label{eq:P_N_approx}
\end{equation}

% It is known that a Poisson distribution $\mathcal{P}(\lambda)$ can be well approximated by a Gaussian distribution $\mathcal{N}(\lambda, \lambda)$,
% when $\lambda$ is sufficiently large (e.g. $\lambda > 10$)~\cite{foi2008practical}. In our application, since $ \frac{x^*} {k}$ is a variable, we plot its distribution in our training dataset in Fig.~\ref{fig:k_sigma}. It suggests that in practice we can treat $ \frac{x^*} {k}$ as a sufficiently large value so that the following approximation holds:

%\begin{equation}
%    P( \frac{x^*} {k}) + \mathcal{N}(\frac{\sigma^2}{k^2},  \frac{\sigma^2}{k^2})
%    \approx \mathcal{P}(\frac{x^*} {k})  + \mathcal{P}(\frac{\sigma^2}{k^2})
%    = \mathcal{P}(\frac{x^*}{k} + \frac{\sigma^2}{k^2}) 
%    = \mathcal{P}[f(x^*)].
%    \label{eq:P_N_approx}
%\end{equation}

Combining Eqn.~\eqref{eq:f_x_dist1} and Eqn.~\eqref{eq:P_N_approx}, the approximate distribution of $f(x)$ is:
\begin{equation}
	f(x) \sim \mathcal{N}[f(x^*), f(x^*)].
\label{eq:noise_k_sigma}
\end{equation}

\begin{figure}[htpb]
	\centering
	\includegraphics[width=.6\linewidth]{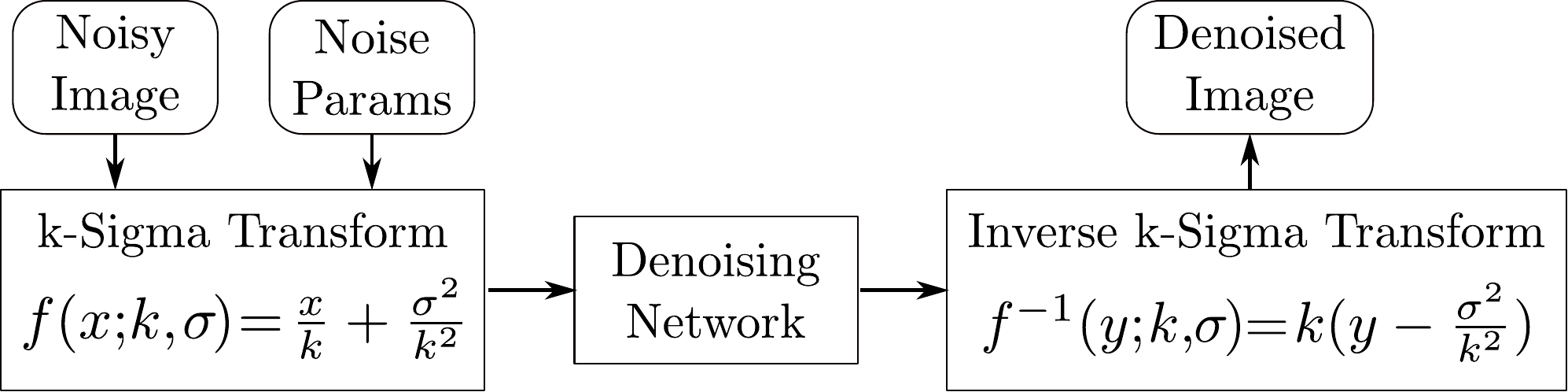}
	\caption{The pipeline of running ISO-independent denoising network with k-Sigma Transform.} %
	\label{fig:denoise_pipeline}
\end{figure}

Eqn.~\eqref{eq:noise_k_sigma} indicates that the distribution of $f(x)$ 
only depends on $f(x^*)$. 
As shown in Fig.~\ref{fig:denoise_pipeline}, we
can train a single network that takes $f(x)$
as input and outputs $f(\hat{x}^*)$ as an estimation of $f(x^*)$. The estimated true
image value $x^*$ can then be computed by applying the inverted k-Sigma Transform 
$f^{-1}(\cdot)$ to $f(\hat{x}^*)$.
In other
words, we apply ISO-dependent transforms to the input and output of the neural
network, so that the network can be trained using normalized data without
considering the ISO setting. 

\section{Learning to Denoise}

%\subsection{Mobile-first Network Architecture}
\subsection{Mobile-friendly Network Architecture}

\begin{figure}[htpb]
    \centering
	\begin{subfigure}[b]{\linewidth}
	\begin{center}
    \includegraphics[width=.9\linewidth]{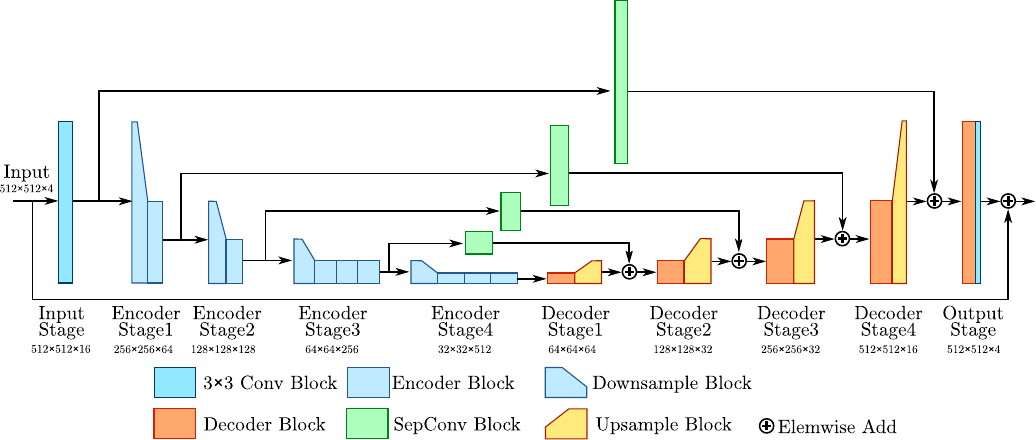}
	\end{center}
	\caption{The U-Net-like overall structure of the denoising network.}
	\label{fig:network:architecture}
	\end{subfigure}

	\begin{subfigure}[b]{\linewidth}
	\begin{center}
    \includegraphics[width=.9\linewidth]{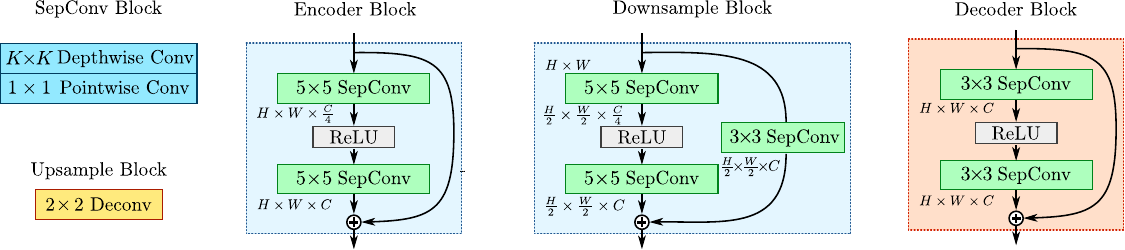}
	\end{center}
	\caption{Detailed stucture of network blocks.}
	\label{fig:network:blocks}
	\end{subfigure}
	\caption{The architecture of the proposed denoising network.}%
	\label{fig:network}
\end{figure}

We further introduce a mobile-friendly convolutional neural network
for image denoising, as shown in Fig.~\ref{fig:network}. We use a 
U-Net-like~\cite{ronneberger2015u} architecture with 4 encoder and 4 decoder stages with skip connections, illustrated in Fig.~\ref{fig:network:architecture}. 
%Fig.~\ref{fig:network} illustrates the architecture of our convolutional neural network
%for image denoising. As shown in Fig.~\ref{fig:network:architecture}, we use a 
%U-Net-like~\cite{ronneberger2015u}
%overall architecture with 4 encoder and 4 decoder stages with skip connections. 

Fig. \ref{fig:network:blocks} depicts the detailed structures of the network blocks. Specifically, in order to run on mobile devices, we use separable-conv \cite{Chollet2016}
in all encoder and decoder stages to reduce the computation cost, and normal dense convolution
layers are only used in the input and output stage.
In encoders, we use $5\times 5$ kernel size to increase receptive field and decrease network depth, 
and downsample feature maps with stride-2 convolutions. 
In decoder, we only use $3\times 3$-speconv and upsample feature maps with $2\times 2$ deconvolutions.  The inputs of each encoder stages are combined into its corresponding decoder stage by element-wise
adding, a $3\times 3$-speconv is adopted in the skip connect to match the channel shape. 
Finally, the last convolution layer outputs a residual added to the input image as the 
denoising result. 

\subsection{Training Dataset}
\label{sec:train:dataset}

To train our denoising network, we need pairs of noisy and clean RAW images. 
In this paper, we use a subset of See-in-the-Dark (SID) dataset proposed in \cite{chen2018learning} as the ground truth clean images. 
The SID dataset contains RAW images captured from a Sony $\alpha$7s~II and a Fujifilm X-T2 camera,
we choose the 10s and 30s long-exposure subset captured by the Sony $\alpha$7s~II camera, and manually
take out those with visible noise, leaving {214} high quality RAW images. %Another method to obtain high quality raw images is unprocessing commonly available Internet photos  described in~\cite{brooks2019unprocessing}.

According to our noise model described in Section \ref{sec:method}, 
if clean RAW images were available, we can synthesize noisy images by sampling from a Poisson-Gaussian
distribution with estimated noise parameters measured from the target sensor. %identical noise parameters to the target sensor.

\subsection{Settings and Details}

To generate tranining samples, we randomly crop $1024\times 1024$-sized bayer patches 
from the original dataset. 
We adopt the bayer-aug method described in~\cite{Liu2019} with random horizontal and 
vertical flipping and ensure the input bayer pattern is in R-G-G-B order.
We then pack the bayer image to $512\times 512\times 4$-shaped RGGB tensor. We also randomly adjust
the brightness and contrast of the cropped images for data augmentation. Noisy
images are then synthesized according to the noise model, with noise parameters of randomly 
selected ISO value. Finally, we apply k-Sigma Transform to both the noisy and clean images so that
the denosing network is trained in the ISO-independent space.

We use $\ell_1$ distance between the noisy and clean images as the loss function, and train 
the network using Adam~\cite{kingma2014adam} optimizer. We adopt the triangular cyclical 
learning rate scheduling~\cite{smith2017cyclical} with the maximum learning rate of 1e-3, 
cycle step of 50$\times$214 iterations, and the base learning rate linearly 
decays to 1e-5 after 4000$\times$214 iterations. 
The batch size is set to 1, and the training converges at 8000$\times$214 iterations.

\section{Experiments}

% datasets and evaluation metrics
In this section, we evaluate our denoising method with a real world dataset collected with an OPPO Reno-10x smartphone. 
This smartphone has three rear cameras, and we use the most commonly used main camera for our test. The sensor of this
camera is Sony IMX586 sized at \SI{1/2}{\arcsecond} with 48 megapixels, and the pixel size is \SI{0.8}{\um}. This sensor
is widely used in many smartphones of 2019, including OPPO Reno series, Xiaomi 9, etc. 

% OPPO Reno-10x uses IMX586's binning mode where the values of adjacent 4-pixels are averaged and outputs 12-megapixel images.
% Therefore, the equavilent pixel size is \SI{1.6}{\um}.

\subsection{Noise Parameters Estimation}

% 验证参数估计的准确度？ 展示数据采集（假）设备（灰度图 + 光照设计）来说明Practical？只是放 K 和 B 的图不够说服力，或者考虑把 噪声参数相关实验 这个section 合并进来

We first measure and estimate the noise parameters with the method described in Section~\ref{sec:method:param}. 
We write a simple app to collect RAW images with Android Camera-2 API, which allows us
manually control the camera's ISO
and exposure time. To keep a stable light condition, we use an Xrite SpectraLight QC light booth in a dark room.
At each ISO and exposure time setting, we adjust the luminance of the light source to avoid over or under exposure,
and the final values of the captured image are kept in an appropriate range.

\begin{figure}[tpb]
	\centering
	\begin{subfigure}[b]{.3\linewidth}
	\begin{center}
		\includegraphics[width=\linewidth]{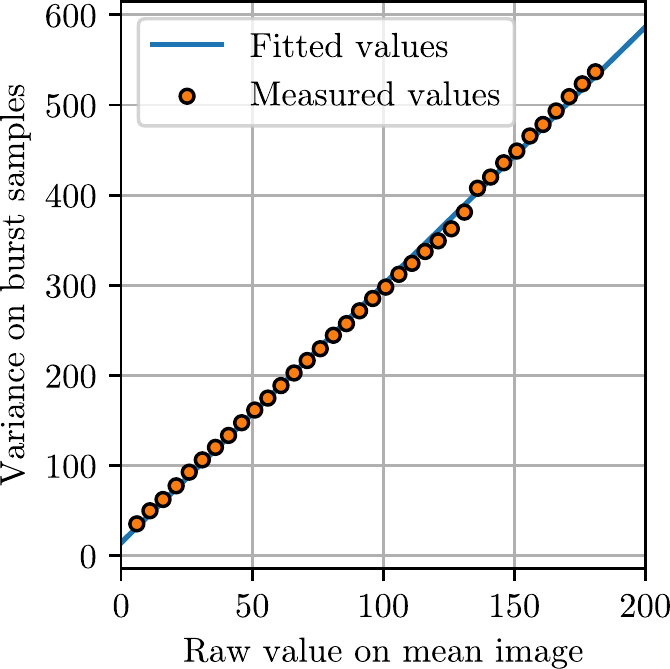}
	\end{center}
	\caption{}%
	\label{fig:noise_iso_4800}
	\end{subfigure}
	\begin{subfigure}[b]{.3\linewidth}
	\begin{center}
		\includegraphics[width=\linewidth]{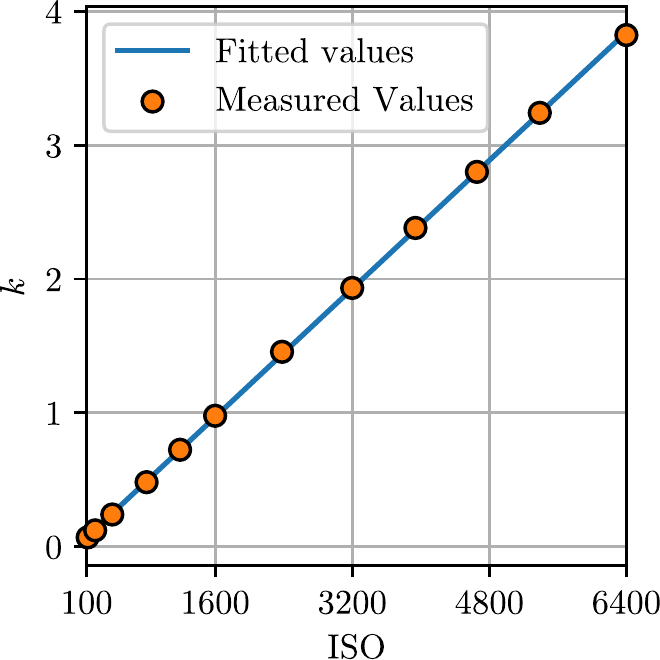}
	\end{center}
	\caption{}
	\label{fig:noise_K}
	\end{subfigure}
	\begin{subfigure}[b]{.3\linewidth}
	\begin{center}
		\includegraphics[width=\linewidth]{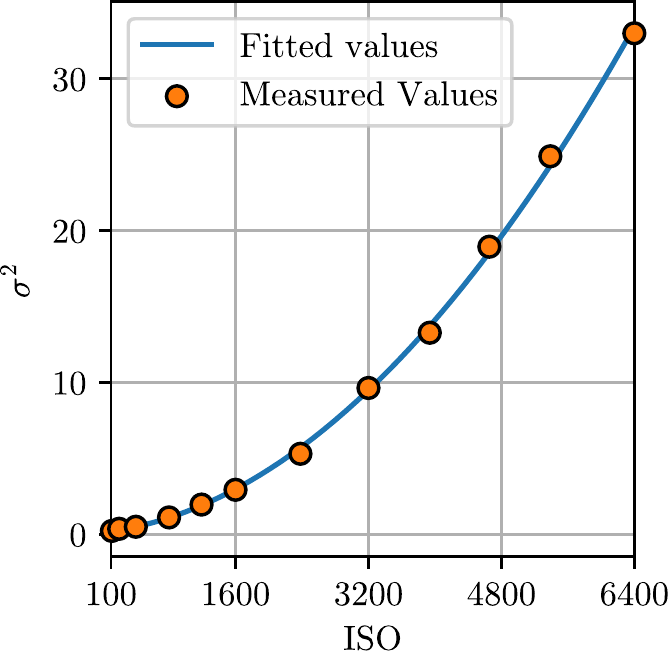}
	\end{center}
	\caption{}
	\label{fig:noise_B}
	\end{subfigure}

	\caption{Noise param estimation of Reno-10x smartphone: %
		(a) parameter estimation at ISO-4800 %
		(b) $k$ values at different ISOs %
		(c) $\sigma^2$ at differnt ISOs.
		}%
	\label{fig:estimated_noise_params}
\end{figure}

At each ISO setting, 64 RAW images are captured in burst mode, and the mean image is considered as 
the clean image. With the method described in Section~\ref{sec:method:param}, we can estimate the 
noise params $k$ and $\sigma^2$ at each specified ISO setting. Fig.~\ref{fig:noise_iso_4800} plots
the value-variance curve of the test phone under ISO \num{4800}, where the scattered dots represent
the measured variances corresponding to each raw value on the mean image, and the blue line plots
the linear regression result of Eqn.~\eqref{eq:p_mean_variance}. From the figure we can see that our
theoretical noise model can well fit the measurement results. 
The slope of the fitted line is the estimated noise parameter $\hat{k}$ and the y-intercept value
is the estimated noise parameter $\hat{\sigma}^2$.

The adjustable range of {IMX586} sensor analog gain is $[1.0, 64.0]$, corresponding to the ISO
value of OPPO Reno-10x camera as $[100, 6400]$.
According to our noise model Eqn.~\eqref{eq:noise_all},
the params $k$ and $\sigma^2$ are linearly, and qudraticly correlated to the ISO value, respectively.
We measure and estimate
the noise params at each ISO setting, and plot the ISO-$k$ and ISO-$\sigma^2$ curve in
Fig.~\ref{fig:estimated_noise_params}. The scattered dots represent the estimated noise params under 
each ISO setting, and the blue lines in Fig.~\ref{fig:noise_K} and Fig.~\ref{fig:noise_B} respectively
represent the linearly and qudraticly fitted curves, which demostrate that our theoretical model matches
the measurements well.

With the ISO-$k$ and ISO-$\sigma^2$ curves well fitted, the noise params under any ISO setting can be 
easily calculated, and thus satisfying the requirements of both synthesizing training data and applying
the k-Sigma transform.

\subsection{Test Dataset and Metrics}

Since our proposed denoising network needs
to be trained for specific sensors, we cannot directly use public benchmarks such as SIDD~\cite{SIDD_2018_CVPR}
due to the mismatching of sensors.
Therefore, we build a testing dataset to evaluate our denoising method. 

\begin{figure}[htpb]
	\centering

	\begin{subfigure}[b]{.24\linewidth}
	\begin{center}
		\includegraphics[width=\linewidth]{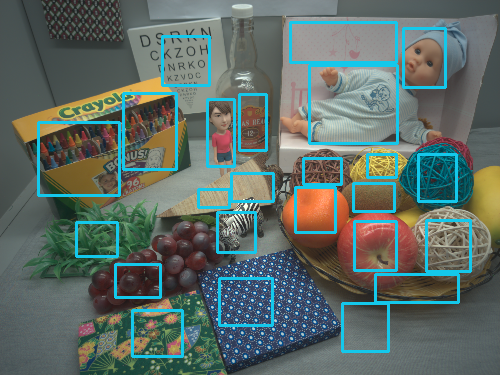}
	\end{center}
	%\caption{Scene 1}
	\end{subfigure}
	\begin{subfigure}[b]{.24\linewidth}
	\begin{center}
		\includegraphics[width=\linewidth]{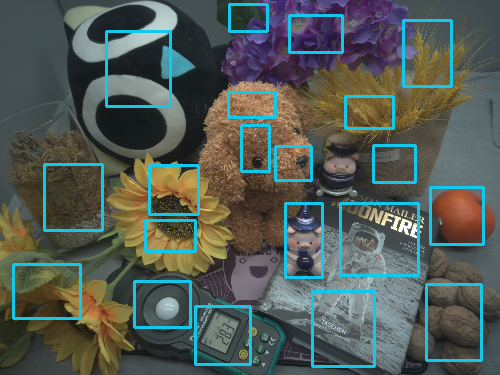}
	\end{center}
	%\caption{Scene 2}
	\end{subfigure}
	\begin{subfigure}[b]{.24\linewidth}
	\begin{center}
		\includegraphics[width=\linewidth]{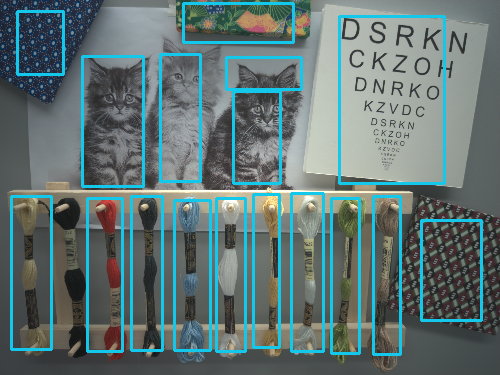}
	\end{center}
	%\caption{Scene 3}
	\end{subfigure}
	\begin{subfigure}[b]{.24\linewidth}
	\begin{center}
		\includegraphics[width=\linewidth]{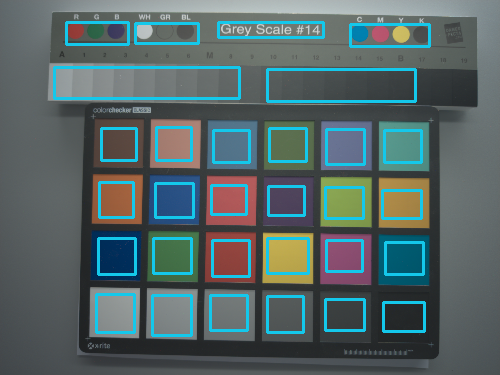}
	\end{center}
	%\caption{Scene 4}
	\end{subfigure}

	\caption{The 4 scenes from the test dataset: the blue boxes represent the regions of interest where image quality metrics
	are calculated.}%
	\label{fig:test_scenes}
\end{figure}

We use an Xrite SpectraLight QC light booth in a dark room to build a stable light condition. 
For each capturing process, 64 RAW images are captured and 
the mean image can be used as the estimated ground truth. 
As shown in Fig.~\ref{fig:test_scenes}, we capture 4 static scenes as the content of testing images, 
and 2 luminance conditions are set 
for each scene. When capturing one scene, the camera position and scene contents are kept
fixed.
We set 5 exposure combinations at each scene and luminance, which are ISO-800@160ms,
ISO-1600@80ms, ISO-3200@40ms, ISO-4800@30ms and ISO-6400@20ms, respectively. These settings
share an identical value of the product of ISO and exposure time, so that the captured images 
have similar brightness but different noise levels.

% The major challenge of building the benchmark dataset for denoising is obtaining the ground truth. 

We use peak signal-noise-ratio~(PSNR) and structural similarity~(SSIM) to measure the performance
of denoising methods. The PSNR and SSIM
between the denoising results and the clean images are measured
in sRGB domain with a simple post processing pipeline, including
\begin{enumerate*}[label=(\arabic*)]
	\item white balance correction,
	\item demosaicking,
	\item color correction, and
	\item gamma correction.
\end{enumerate*}
The parameters for white balance correction and color correction are obtained from the metadata
of the RAW image. The demosaicking algorithm is pixel-grouping~(PPG) and the gamma value is
\num{2.2}.

\subsection{Results}

We first show the comparison between our method and the previous state-of-the-art raw image denoising 
method proposed in~\cite{Liu2019}, which is trained and tested using the SIDD dataset. An extra-large
UNet-like network which costs 1T multiply-and-cumulate operations~(MACs) per megapixel is proposed in~\cite{Liu2019}
and achieved the state-of-the-art performance in NTIRE~2019 denoising challenge. In addition to the 
UNet-1T architecture, we modify the network by reducing the channel width and layer depth
to fit various computation complexities.
We train these models with two different data sources: 
SIDD dataset and the training data generated by our method. 

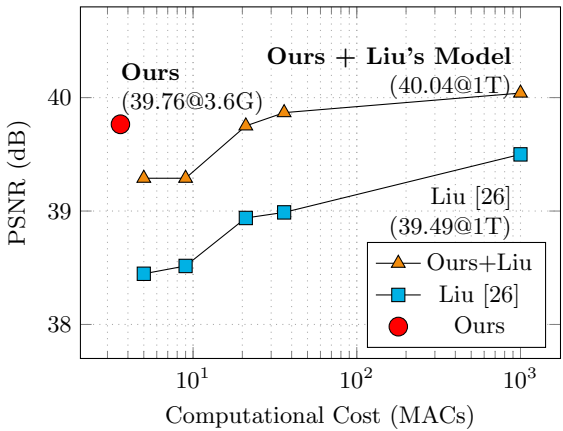
\begin{figure}[tb]
	\centering
	\begin{tikzpicture}

\definecolor{mr-grid}{HTML}{AAAAAA}
\definecolor{mr-SIDD}{HTML}{00B1E4}
\definecolor{mr-Ours}{HTML}{F28C0D}

\pgfplotsset{compat=1.8}
\pgfplotsset{every tick label/.append style={font=\footnotesize}}
\pgfplotsset{grid style={dotted,mr-grid}}

\begin{axis}[
		xmode=log,
		ymin=37.7, ymax=40.8,
		grid=both,
		legend pos=south east,
		width=8cm,
		height=6.2cm,
		xlabel={\small Computational Cost~(MACs)},
		ylabel={\small PSNR~(\si{\decibel})}
]
	\addplot [mark size=3pt, mark=triangle*, mark options = {draw=black, fill=mr-Ours}]
	table {%
	x	y
	% 3.6 39.7648896
	5.0 39.289135
	9.0 39.289135
	21.0 39.750575
	36.0 39.867471
	1000 40.04
	};
	
	\addplot [mark size=2.5pt, mark=square*, mark options = {draw=black, fill=mr-SIDD}]
	table {%
	x	y
	% 3.6 39.21411855
	5.0 38.446864
	9.0  38.51576
	21.0 38.93888553
	36.0  38.98807406
	1000 39.49952
	};
	
	\addplot[only marks, draw=black, fill=red, mark size=3.5pt, mark=*]
	table {%
	x	y
	3.6 39.7648896
	};
	
	\legend{Ours+Liu, Liu~\cite{Liu2019}, Ours}
	
	\node at (axis cs:3.2,39.9)[
		anchor=base west,
		text=black,
		rotate=0.0,
		align=left
	]{\bf\small Ours\\(39.76@3.6G)};
	
	\node at (axis cs:1000,40.04)[
		anchor=base east,
		text=black,
		rotate=0.0,
		align=right
	]{\bf\small Ours + Liu's Model\\(40.04@1T)};

	\node at (axis cs:1000,38.8)[
		anchor=base east,
		text=black,
		rotate=0.0,
		align=right
	]{Liu~\cite{Liu2019}\\(39.49@1T)};

% \addplot [only marks, draw=black, fill=blue, mark size=3pt]
% table {%
% x     y
% +7.14 +0.6085
% +1.61 +0.6477
% +0.476 +0.5358
% };
% \addplot [only marks, draw=black, fill=red, mark size=3pt]
% table {%
% x           y
% +13.245e+00 +0.7820
% +16.779e+00 +0.7571
% };
% 
% \node at (axis cs:7.3,0.58)[
%   anchor=base west,
%   text=black,
%   rotate=0.0,
%   align=left
% ]{Tian~\cite{tian2016detecting}\\(0.609@7.14fps)};
% \node at (axis cs:1.3,0.6677)[
%   anchor=base west,
%   text=black,
%   rotate=0.0,
%   align=left
% ]{Yao~\cite{yao2016scene}\\(0.648@1.61fps)};
% 
% \node at (axis cs:0.6,0.505)[
%   anchor=base west,
%   text=black,
%   rotate=0.0,
%   align=left
% ]{Zhang~\cite{Ref:Zhang2016}\\(0.532@0.476fps)};
% 
% \node at (axis cs:13.0,0.78)[
%   anchor=base east,
%   text=black,
%   rotate=0.0,
%   align=right
% ]{\bf Ours+PVANet2x\\(0.782@13.2fps)};
% 
% \node at (axis cs:16.8,0.69)[
%   anchor=base east,
%   text=black,
%   rotate=0.0,
%   align=right
% ]{\bf Ours+PVANet\\(0.757@16.8fps)};
\end{axis}

\end{tikzpicture}

% vim: ft=tex
	\caption{PSNRs at different computation costs of our method, our method using Liu's model~\cite{Liu2019}, and Liu's method. Note that the MACs are in the log space.}%
	\label{fig:flops_psnr}
\end{figure}

\begin{figure}[tpb]
	\centering
	\includegraphics[width=0.9\linewidth]{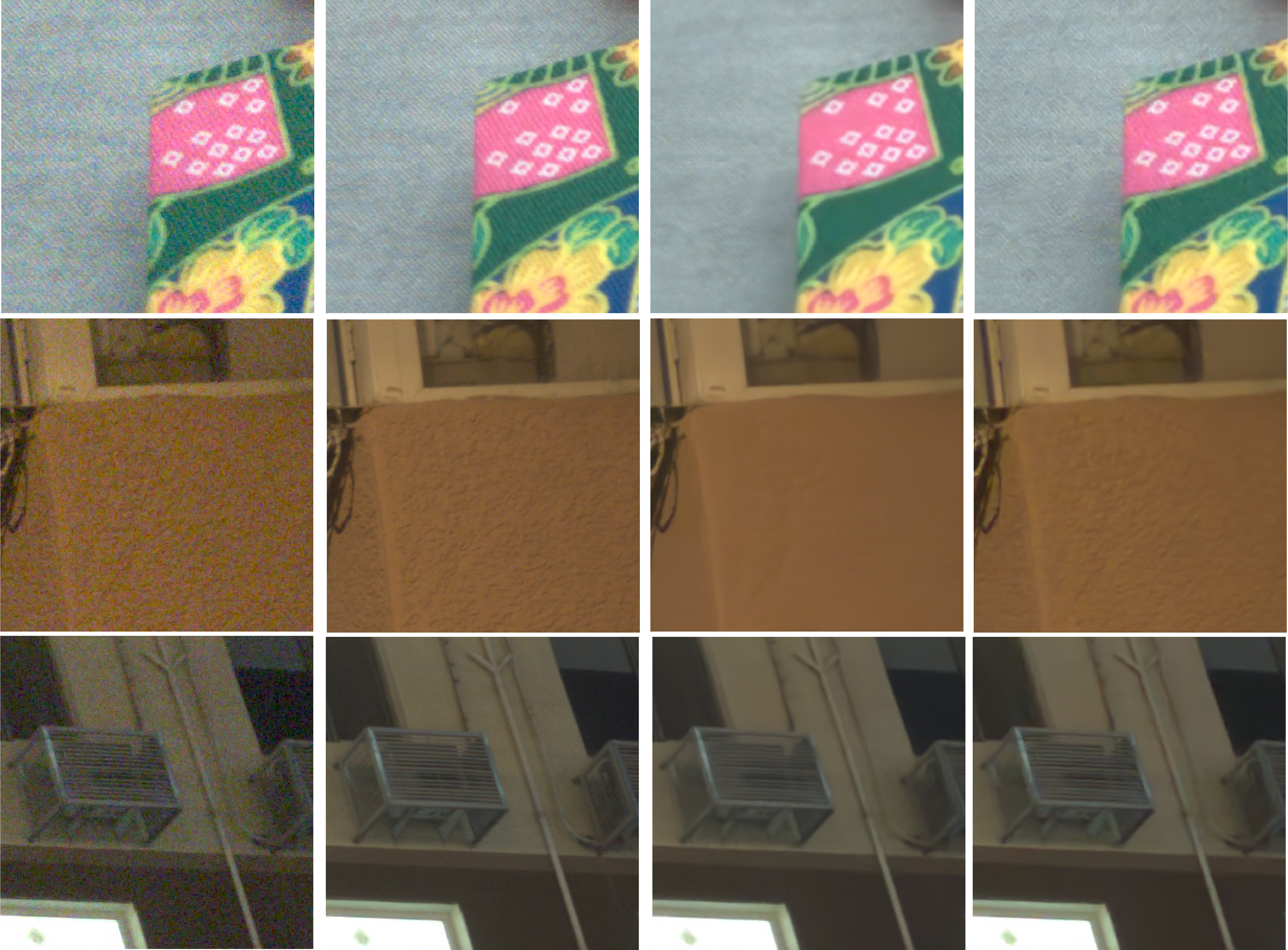}
	\caption{More visual results on our real dataset. From left to right: input image; ground truth; result based on method in  \cite{Liu2019}; our result. Compared with \cite{Liu2019} which generates blurred areas, our method can efficiently reduce the noise as well maintain underlying details. Moreover, our method utilizes a significantly smaller network (3.6G vs 1T).}%
	\label{fig:more_results}
\end{figure}

As shown in Fig.~\ref{fig:flops_psnr}, 
because of our accurate noise modeling, the models trained with our synthetic data outperform 
those trained with the SIDD dataset by a large margin. Moreover, our mobile-friendly network 
trained on our synthetic data achieves comparable performance with the UNet-36G 
with only $10\%$ of its computational complexity (3.6G vs 36.3G). 
More visual comparisons are provided in Fig.~\ref{fig:more_results}.

\begin{table}[tb]
	\caption{Running time of denoising 1~MP on Qualcomm Snapdragon 855 GPU}
	\label{tab:speed_test}
	\centering
	\begin{tabular}{lrrrrr}
		\toprule
		Method & Ours & UNet-5G & UNet-21G & UNet-36G \\
		\midrule
		\si{\ms}/MP & 70.70 & 79.19 & 292.0 & 383.9 \\
		\bottomrule
	\end{tabular}
\end{table}

We further test the actual running time of different models on mobile devices, listed in Table~\ref{tab:speed_test}. Our mobile-friendly model can process a $1024 \times 1024$ Bayer input with \SI{70.70}{\ms} (aka $\sim$\SI{850}{\ms} for a 12MP full-sized image) on a Qualcomm Snapdragon 855 GPU, while other models with comparable performance require significantly longer time, \SI{292}{\ms} for 21G network and \SI{383}{\ms} for 36G network (aka \SI{4.45}{\s} for 12MP), making them impractical to be deployed on mobile devices.

\subsection{Ablation Studies}
\subsubsection{Data synthesis method}
\sisetup{round-mode=figures, round-precision=4}

To verify the effectiveness of our data synthesis method, we train our denosing network with four different  training datasets, including
\begin{itemize}
\item testset overfitting: directly use the inputs and ground truth from the testset as the training data;
\item testset with synthetic: use the ground truth of the testset and add synthetic noise as the input;
\item testset with noise scaling: use our noise synthesis method but scale the noise parameters to purposely mismatch with the target sensor;
\item SID: synthesize the training set with SID dataset described in Section~\ref{sec:train:dataset}.
\end{itemize}

The test results are listed in Table~\ref{tab:noise_param_test}. Not surprisingly, the overfitting experiment obtains
the highest PSNR of \SI{40.06}{\decibel} and SSIM of \num{0.9335}, which sets an upper-bound for this comparison. 
The testset with our synthetic noise achieves the second best result of \SI{39.97}{\decibel} PSNR.
When using
inaccurate noise parameters by scaling $k$ and $\sigma^2$ by 0.5 or 2.0, the network results in the lowest
PSNRs in this experiment. Our method of using the SID dataset achieves 
\SI{39.76}{\decibel} PSNR.

This experiment shows that our noise model and synthetic noise can well match the characteristics of 
the real input noise, and the testing results can be close to the upper bound performance. 
Inaccurate noise parameters, even in the overfitting experiment, lead to noticeable performance degradation.

\begin{table}[tpb]
	\centering
	\caption{Comparison of training datasets and noise parameters}
	\label{tab:noise_param_test}
	\begin{tabular}{lS[detect-weight]S[detect-weight]}
		\toprule
		Dataset & {PSNR~(\si{\decibel})} & SSIM \\
		\midrule
		testset overfitting & 40.05706288 & 0.933505932 \\
		testset+synthetic & 39.96560621 & 0.933691115 \\
		testset+synthetic scale 0.5 & 36.87384014 & 0.885018911  \\
		testset+synthetic scale 2.0 & 39.68042145 & 0.93010538 \\
		% SIDD & 38.1597996 & 0.916650591 \\
		SID & 39.76488957 & 0.931048368 \\
		\bottomrule
	\end{tabular}
\end{table}

\subsubsection{Robustness to ISO}
\sisetup{round-mode=places, round-precision=2, detect-weight=true}

We compare several strategies of denoising images of different ISO settings to verify the effectiveness of the k-Sigma Transform.
We compare two ways of handling multiple ISO settings: 
\begin{enumerate*}[label=(\arabic*)]
    \item iso-augmentation: randomly choose ISO settings when synthesizing training samples and directly feed them into the 
    denosing network; and
    \item concat-variance: the method proposed in~\cite{brooks2019unprocessing}, where the estimated noise variance 
    are concatenated to the input as 4 additional channels.
\end{enumerate*}
In addition, we also test the performance of the single-ISO method, where the training data is synthesized using noise
parameters of a single ISO.

The results are listed in Table~\ref{tab:multi_iso}. The concat-variance strategy achieves
the PSNR of \SI{39.65}{\decibel}, which is about \SI{0.09}{\decibel} higher than the iso-augmentation strategy.
This means that explicit noise level information can help the model achieve better results than blind denoising.
With the proposed k-Sigma Transform, our network achieves the highest PSNR
in this experiment. In comparison, all single-ISO methods perform much worse than multi-ISO ones.

\sisetup{round-mode=figures, round-precision=4, detect-weight=true}
\begin{table}[tpb]
	\centering
	\caption{Strategies of denoising for multiple ISOs}
	\label{tab:multi_iso}
	\begin{tabular}{lSS}
		\toprule
		Method & {PSNR~(\si{\decibel})} & SSIM \\
		\midrule
		k-Sigma Transform & 39.76 & 0.931048368 \\
		concat-variance~\cite{brooks2019unprocessing} & 39.653042266876 & 0.930745060240976 \\
		iso-augmentation & 39.56662978 & 0.929922157 \\
		single-iso-1600 & 35.73569171524029 & 0.8072025198136967 \\
		single-iso-3200 & 38.561628588767086 & 0.9089081792578051 \\
		single-iso-6400 & 38.1597996 & 0.916650591 \\
		\bottomrule
	\end{tabular}
\end{table}

Table~\ref{tab:different_iso} gives a more detailed analysis of single-ISO methods and our approach, where
the PSNRs measured in the testset are grouped into different ISO settings. From the table we can see that when
the ISO setting of the testset matches with the single-ISO model, it can produce competitive denoising results.
Our method base on the k-Sigma Transform performs consistently well under all ISO settings.

\begin{table}[tpb]
	\centering
	\caption{PSNRs under different ISO settings}
	\label{tab:different_iso}
	\begin{tabular}{lccccc}
	\toprule
	  & ISO-800 & ISO-1600 & ISO-3200 & ISO-4800 & ISO-6400 \\
	\midrule
	k-Sigma Transform & 43.21 & \textbf{41.48} & \textbf{39.49} & 38.17 & \textbf{36.94} \\
	% iso-aug & 42.99 & 41.32 & 39.33 & 37.92 & 36.73 \\
	single-iso-1600 & 42.96 & \textbf{41.48} & 35.01 & 31.33 & 28.86 \\
	single-iso-3200 & 41.79 & 40.87 & \textbf{39.51} & 36.97 & 34.08 \\
	single-iso-6400 & 39.59 & 38.59 & 38.11 & 37.80 & \textbf{36.91} \\
	\bottomrule
	\end{tabular}
\end{table}

%\subsection{Application}

%\subsection{Ablation Study}
%\subsubsection{Noise-Level Robustness}
%\subsubsection{Model-size Comparison}
%\subsection{Speed on Mobile Device}

\section{Conclusion}
We have presented a new raw image denoiser designed for mobile devices. By accurate sensor noise estimation, we can utilize a light-weight network trained on sensor-specific synthetic data that generalizes well to real noise. We also propose a k-Sigma Transform to process the input and output data, so that denoising can be learned in an ISO-independent space. This allows the network to handle a wide range of noise level without increasing network complexity. Our results show that the proposed method can achieve compatible performance with state-of-the-art methods, which typically employ much larger networks that cannot be directly applied for mobile applications. 

In applications, our method can be integrated into existing camera pipeline and replace its denoising component. Since our method can produce high quality denoised raw images, it gives a strong base for the ISP to apply more aggressive post-processing. Our methods have been featured in the night shot mode of several flagship phones released in 2019, with stable and outstanding performance on mobile devices. 

In the future, we would like to explore how to further reduce the computational complexity of the proposed method, so that we can apply it on video streams in real-time. Also, we believe it will be interesting and promising to explore deep learning based approaches for raw image processing that can improve or even replace camera's ISP pipeline.   

%Our method can process the raw images directly from sensor and pass it to ISP for additional enhancement.

%\input{todo}

\clearpage

\bibliographystyle{splncs04}
\bibliography{egbib}

\begin{thebibliography}{10}
\providecommand{\url}[1]{\texttt{#1}}
\providecommand{\urlprefix}{URL }
\providecommand{\doi}[1]{https://doi.org/#1}

\bibitem{SIDD_2018_CVPR}
Abdelhamed, A., Lin, S., Brown, M.S.: A high-quality denoising dataset for
  smartphone cameras. In: The IEEE Conference on Computer Vision and Pattern
  Recognition (CVPR) (June 2018)

\bibitem{aharon2006k}
Aharon, M., Elad, M., Bruckstein, A., et~al.: K-svd: An algorithm for designing
  overcomplete dictionaries for sparse representation. IEEE Transactions on
  signal processing  \textbf{54}(11), ~4311 (2006)

\bibitem{anaya2018renoir}
Anaya, J., Barbu, A.: Renoir--a dataset for real low-light image noise
  reduction. Journal of Visual Communication and Image Representation
  \textbf{51},  144--154 (2018)

\bibitem{anscombe1948transformation}
Anscombe, F.J.: The transformation of poisson, binomial and negative-binomial
  data. Biometrika  \textbf{35}(3/4),  246--254 (1948)

\bibitem{brooks2019unprocessing}
Brooks, T., Mildenhall, B., Xue, T., Chen, J., Sharlet, D., Barron, J.T.:
  Unprocessing images for learned raw denoising. In: Proceedings of the IEEE
  Conference on Computer Vision and Pattern Recognition. pp. 11036--11045
  (2019)

\bibitem{buades2005non}
Buades, A., Coll, B., Morel, J.M.: A non-local algorithm for image denoising.
  In: 2005 IEEE Computer Society Conference on Computer Vision and Pattern
  Recognition (CVPR'05). vol.~2, pp. 60--65. IEEE (2005)

\bibitem{burger2012image}
Burger, H.C., Schuler, C.J., Harmeling, S.: Image denoising: Can plain neural
  networks compete with bm3d? In: CVPR (2012)

\bibitem{chen2018learning}
Chen, C., Chen, Q., Xu, J., Koltun, V.: Learning to see in the dark. In:
  Proceedings of the IEEE Conference on Computer Vision and Pattern
  Recognition. pp. 3291--3300 (2018)

\bibitem{chen2018image}
Chen, J., Chen, J., Chao, H., Yang, M.: Image blind denoising with generative
  adversarial network based noise modeling. In: CVPR (2018)

\bibitem{chen2017trainable}
Chen, Y., Pock, T.: Trainable nonlinear reaction diffusion: A flexible
  framework for fast and effective image restoration. IEEE transactions on
  pattern analysis and machine intelligence  \textbf{39}(6),  1256--1272 (2017)

\bibitem{Chollet2016}
Chollet, F.: {Xception: Deep Learning with Depthwise Separable Convolutions}
  (oct 2016), \url{http://arxiv.org/abs/1610.02357}

\bibitem{dabov2008image}
Dabov, K., Foi, A., Katkovnik, V., Egiazarian, K.: Image restoration by sparse
  3d transform-domain collaborative filtering. In: Image Processing: Algorithms
  and Systems VI. vol.~6812, p. 681207. International Society for Optics and
  Photonics (2008)

\bibitem{elad2006image}
Elad, M., Aharon, M.: Image denoising via sparse and redundant representations
  over learned dictionaries. IEEE Transactions on Image processing
  \textbf{15}(12),  3736--3745 (2006)

\bibitem{EMVA2010}
{European Machine Vision Association.}: {Standard for Characterization of Image
  Sensors and Cameras} (2010). \doi{10.1063/1.1518010}

\bibitem{foi2007noise}
Foi, A., Alenius, S., Katkovnik, V., Egiazarian, K.: Noise measurement for
  raw-data of digital imaging sensors by automatic segmentation of nonuniform
  targets. IEEE Sensors Journal  \textbf{7}(10),  1456--1461 (2007)

\bibitem{foi2008practical}
Foi, A., Trimeche, M., Katkovnik, V., Egiazarian, K.: Practical
  poissonian-gaussian noise modeling and fitting for single-image raw-data.
  IEEE Transactions on Image Processing  \textbf{17}(10),  1737--1754 (2008)

\bibitem{gharbi2016deep}
Gharbi, M., Chaurasia, G., Paris, S., Durand, F.: Deep joint demosaicking and
  denoising. ACM Transactions on Graphics (TOG)  \textbf{35}(6), ~191 (2016)

\bibitem{gu2014weighted}
Gu, S., Zhang, L., Zuo, W., Feng, X.: Weighted nuclear norm minimization with
  application to image denoising. In: CVPR (2014)

\bibitem{Hasinoff2016}
Hasinoff, S.W., Sharlet, D., Geiss, R., Adams, A., Barron, J.T., Kainz, F.,
  Chen, J., Levoy, M.: {Burst photography for high dynamic range and low-light
  imaging on mobile cameras}. ACM Transactions on Graphics  \textbf{35}(6),
  1--12 (2016). \doi{10.1145/2980179.2980254},
  \url{http://dl.acm.org/citation.cfm?doid=2980179.2980254}

\bibitem{hirakawa2006joint}
Hirakawa, K., Parks, T.W.: Joint demosaicing and denoising. IEEE Transactions
  on Image Processing  \textbf{15}(8),  2146--2157 (2006)

\bibitem{jain2009natural}
Jain, V., Seung, S.: Natural image denoising with convolutional networks. In:
  Advances in neural information processing systems. pp. 769--776 (2009)

\bibitem{kingma2014adam}
Kingma, D.P., Ba, J.: Adam: A method for stochastic optimization. arXiv
  preprint arXiv:1412.6980  (2014)

\bibitem{lehtinen2018noise2noise}
Lehtinen, J., Munkberg, J., Hasselgren, J., Laine, S., Karras, T., Aittala, M.,
  Aila, T.: Noise2noise: Learning image restoration without clean data. arXiv
  preprint arXiv:1803.04189  (2018)

\bibitem{Liba2019}
Liba, O., Murthy, K., Tsai, Y.T., Brooks, T., Xue, T., Karnad, N., He, Q.,
  Barron, J.T., Sharlet, D., Geiss, R., Hasinoff, S.W., Pritch, Y., Levoy, M.:
  {Handheld mobile photography in very low light}. ACM Transactions on Graphics
   \textbf{38}(6) (2019). \doi{10.1145/3355089.3356508}

\bibitem{liu2008automatic}
Liu, C., Szeliski, R., Kang, S.B., Zitnick, C.L., Freeman, W.T.: Automatic
  estimation and removal of noise from a single image. IEEE Trans. Pattern
  Anal. Mach. Intell.  \textbf{30}(2),  299--314 (2008)

\bibitem{Liu2019}
Liu, J., Wu, C.H., Wang, Y., Xu, Q., Zhou, Y., Huang, H., Wang, C., Cai, S.,
  Ding, Y., Fan, H., Wang, J.: {Learning Raw Image Denoising with Bayer Pattern
  Unification and Bayer Preserving Augmentation}  (apr 2019),
  \url{http://arxiv.org/abs/1904.12945}

\bibitem{liu2014practical}
Liu, X., Tanaka, M., Okutomi, M.: Practical signal-dependent noise parameter
  estimation from a single noisy image. IEEE Transactions on Image Processing
  \textbf{23}(10),  4361--4371 (2014)

\bibitem{mairal2009non}
Mairal, J., Bach, F.R., Ponce, J., Sapiro, G., Zisserman, A.: Non-local sparse
  models for image restoration. In: ICCV. vol.~29, pp. 54--62. Citeseer (2009)

\bibitem{makitalo2010optimal}
Makitalo, M., Foi, A.: Optimal inversion of the anscombe transformation in
  low-count poisson image denoising. IEEE transactions on Image Processing
  \textbf{20}(1),  99--109 (2010)

\bibitem{mao2016image}
Mao, X., Shen, C., Yang, Y.B.: Image restoration using very deep convolutional
  encoder-decoder networks with symmetric skip connections. In: NeurIPS (2016)

\bibitem{Mildenhall2017}
Mildenhall, B., Barron, J.T., Chen, J., Sharlet, D., Ng, R., Carroll, R.:
  {Burst Denoising with Kernel Prediction Networks}  (dec 2017),
  \url{https://arxiv.org/abs/1712.02327}

\bibitem{portilla2003image}
Portilla, J., Strela, V., Wainwright, M.J., Simoncelli, E.P.: Image denoising
  using scale mixtures of gaussians in the wavelet domain. IEEE Trans Image
  Processing  \textbf{12}(11) (2003)

\bibitem{ronneberger2015u}
Ronneberger, O., Fischer, P., Brox, T.: U-net: Convolutional networks for
  biomedical image segmentation. In: International Conference on Medical image
  computing and computer-assisted intervention. pp. 234--241. Springer (2015)

\bibitem{guo2018toward}
Shi, G., Zifei, Y., Kai, Z., Wangmeng, Z., Lei, Z.: Toward convolutional blind
  denoising of real photographs. In: arXiv preprint arXiv:1807.04686 (2018)

\bibitem{smith2017cyclical}
Smith, L.N.: Cyclical learning rates for training neural networks. In: 2017
  IEEE Winter Conference on Applications of Computer Vision (WACV). pp.
  464--472. IEEE (2017)

\bibitem{tai2017memnet}
Tai, Y., Yang, J., Liu, X., Xu, C.: Memnet: A persistent memory network for
  image restoration. In: Proceedings of the IEEE international conference on
  computer vision. pp. 4539--4547 (2017)

\bibitem{ulyanov2018deep}
Ulyanov, D., Vedaldi, A., Lempitsky, V.: Deep image prior. In: Proceedings of
  the IEEE Conference on Computer Vision and Pattern Recognition. pp.
  9446--9454 (2018)

\bibitem{xie2012image}
Xie, J., Xu, L., Chen, E.: Image denoising and inpainting with deep neural
  networks. In: Advances in neural information processing systems. pp. 341--349
  (2012)

\bibitem{xu2017multi}
Xu, J., Zhang, L., Zhang, D., Feng, X.: Multi-channel weighted nuclear norm
  minimization for real color image denoising. In: ICCV (2017)

\bibitem{yair2018multi}
Yair, N., Michaeli, T.: Multi-scale weighted nuclear norm image restoration.
  In: CVPR (2018)

\bibitem{zhang2018ffdnet}
Zhang, K., Zuo, W., Zhang, L.: Ffdnet: Toward a fast and flexible solution for
  cnn based image denoising. IEEE Transactions on Image Processing  (2018)

\bibitem{zhou2019awgn}
Zhou, Y., Jiao, J., Huang, H., Wang, Y., Wang, J., Shi, H., Huang, T.: When
  awgn-based denoiser meets real noises. arXiv preprint arXiv:1904.03485
  (2019)

\bibitem{zhou2018survey}
Zhou, Y., Liu, D., Huang, T.: Survey of face detection on low-quality images.
  In: 2018 13th IEEE International Conference on Automatic Face \& Gesture
  Recognition (FG 2018). pp. 769--773. IEEE (2018)

\bibitem{zhu2016noise}
Zhu, F., Chen, G., Heng, P.A.: From noise modeling to blind image denoising.
  In: CVPR (2016)

\end{thebibliography}
\end{document}